\documentclass[twocolumn,showpacs,superscriptaddress,preprintnumbers,amsmath,amssymb]{revtex4}
\usepackage{graphicx}
\usepackage{dcolumn}
\usepackage{bm}

\def\simge{\mathrel{
     \rlap{\raise 0.511ex \hbox{$>$}}{\lower 0.511ex \hbox{$\sim$}}}}
\def\simle{\mathrel{
     \rlap{\raise 0.511ex \hbox{$<$}}{\lower 0.511ex \hbox{$\sim$}}}}

\begin{document}
\preprint{SPhT-T07/028}
\title{The centrality dependence of elliptic flow, the hydrodynamic
  limit, and the viscosity of hot QCD}

\author{Hans-Joachim Drescher}
\affiliation{
Frankfurt Institute for Advanced Studies (FIAS),
Johann Wolfgang Goethe-Universit\"at,
Max-von-Laue-Str.~1, 60438  Frankfurt am Main, Germany
}
\author{Adrian Dumitru}
\affiliation{
Institut f\"ur Theoretische Physik,
Johann Wolfgang Goethe-Universit\"at,
Max-von-Laue-Str. 1, 60438  Frankfurt am Main, Germany
}
\author{Cl\'ement Gombeaud}
\author{Jean-Yves Ollitrault}
\affiliation{Service de Physique Th\'eorique, CEA/DSM/SPhT,
  CNRS/MPPU/URA2306\\ CEA Saclay, F-91191 Gif-sur-Yvette Cedex.}

\date{\today}
\begin{abstract}
We show that the centrality and system-size dependence of elliptic
flow measured at RHIC are fully described by a simple model based on
eccentricity scaling and incomplete thermalization.  We argue that the
elliptic flow is at least 25\% below the (ideal) ``hydrodynamic
limit'', even for the most central Au-Au collisions. This lack of
perfect equilibration allows for estimates of the effective parton
cross section in the Quark-Gluon Plasma and of its viscosity to
entropy density ratio. We also show how the initial conditions affect the 
transport coefficients and thermodynamic quantities extracted from the
data, in particular the viscosity and the speed of sound.
\end{abstract}
\maketitle

When two ultrarelativistic nuclei collide at non-zero impact
parameter, their overlap area in the transverse plane has a short axis,
parallel to the impact parameter, and a long axis perpendicular to it. 
This almond shape of the initial profile is converted by the pressure gradient
into a momentum asymmetry,
so that more particles are emitted along the short 
axis~\cite{Ollitrault:1992bk}. 
The magnitude of this effect is characterized by elliptic flow,
defined as 
\begin{equation}
\label{defv2}
v_2\equiv\langle\cos 2(\varphi-\Phi_R)\rangle,
\end{equation}
where $\varphi$ is the azimuthal angle of an outgoing particle, $\Phi_R$ is
the azimuthal angle of the impact parameter,  and 
angular brackets denote an average over many particles and many 
events. The unexpected large magnitude of elliptic flow at
RHIC~\cite{Ackermann:2000tr} has generated a lot of
activity in recent years. 

Elliptic flow results from the interactions between the produced
particles, and can be used to probe local thermodynamic equilibrium. 
If the produced matter equilibrates, it behaves as an ideal fluid. 
Hydrodynamics predicts that at a given energy, $v_2$ 
scales like the eccentricity $\varepsilon$ of the 
almond~\cite{Ollitrault:1992bk,Sorge:1998mk}. 
It is independent of its transverse size $R$, as a
consequence of the scale invariance of ideal-fluid dynamics. 
If, on the other hand, equilibration is incomplete, then
eccentricity scaling is broken and $v_2/\varepsilon$ also depends on
the Knudsen number $K=\lambda/R$, where $\lambda$ is the length scale
over which a parton is deflected by a large angle.

Here, we show that the centrality dependence of $v_2/\varepsilon$, for
both Au+Au and Cu+Cu collisions, can be described by the following simple
formula~\cite{Bhalerao:2005mm}: 
\begin{equation}
\label{v2k}
\frac{v_2}{\varepsilon}=\frac{v_2^{\rm hydro}}{\varepsilon}\frac{1}{1+K/K_0}~.
\end{equation}
$v_2/\varepsilon$ is largest in the hydrodynamic limit $K\to 0$.  The
first order corrections to this limit, corresponding to viscous
effects, are linear in $K$. For large mean-free path, far from the
hydrodynamic limit, $v_2/\varepsilon\sim1/K$ vanishes like the number
of collisions per particle.  One expects the transition between these
two regimes to occur when $\lambda\simeq R$, hence that $K_0\simeq1$.
A recent transport calculation~\cite{Gombeaud:2007ub} in two spatial
dimensions indeed obtained $K_0\simeq 0.7$.

Elliptic flow develops gradually during the early stages of the collision.
Due to the strong longitudinal expansion, the thermodynamic 
properties of the medium depend on the time $\tau$, of course. The
average particle density, for instance, decreases like $1/\tau$ (if
their number is approximately conserved, see recent discussion
in~\cite{DMN}):
\begin{equation}  \label{density}
\rho(\tau) = \frac{1}{\tau S} \frac{dN}{dy},
\end{equation}
where $dN/dy$ denotes the total (charged + neutral) multiplicity per unit
rapidity, and $S$ is the transverse overlap area between the two nuclei.
The quantities that we shall extract from $v_2$ should
be intepreted as averages over the transverse area $S$, and over 
some time interval around $R/c_s$, which is the typical time scale for the
build-up of $v_2$ in hydrodynamics~\cite{Bhalerao:2005mm}. $c_s$ denotes
the velocity of sound. 

The Knudsen number $K$ is defined by evaluating the mean free path
$\lambda=1/\sigma\rho$ ($\sigma$ is a partonic cross section) at
$\tau=R/c_s$. Thus,
\begin{equation}
\label{knud}
\frac{1}{K} 
=\frac{\sigma}{S}\frac{dN}{dy} \, c_s~.
\end{equation}
The purpose of this Letter is to show that the centrality and
system-size dependence of the data for $v_2$ at RHIC is described very
well by Eqs.~(\ref{v2k}) and~(\ref{knud}). This provides three
important pieces of information. First, such a fit allows us to
``measure'' the Knudsen number corresponding to a given centrality,
which quantifies how close the dense matter produced in heavy-ion
collisions at RHIC is to perfect fluidity. Second, the extrapolation
to $K=0$ allows us to read off the limiting value for $v_2^{\rm
hydro}/\varepsilon$ extracted from the {\em data}; this is useful for
constraining the equation of state (EoS) of QCD via hydrodynamic
simulations, and we shall also see that it exhibits a rather
surprising dependence on the initial conditions. Finally, using 
Eq.~(\ref{knud}), 
we can convert the Knudsen number into the typical
partonic cross section $\sigma$ (and viscosity) in the Quark-Gluon
Plasma (QGP). 
Since only the combination $K_0 \sigma c_s$ 
actually appears in Eq.~(\ref{v2k}), uncertainties in $K_0$ or $c_s$ 
then translate into corresponding uncertainties of $\sigma$. 
Unless mentioned otherwise, our standard choice is 
$c_s=1/\sqrt{3}\simeq 0.58$ (ideal quark-gluon plasma) and $K_0= 0.7$.  
Letting $K_0=1$ and $c_s^2=2/3$\footnote{Such a ``hard'' EoS can arise from
  repulsive long-range interactions among the
  partons such as classical fields. We thank V.~Koch for pointing this
out to us.} instead reduces the 
estimated $\sigma$ by a factor of two; on the other hand, taking
$K_0=0.5$ and $c_s^2=1/6$ increases $\sigma$ by the same factor.

For the elliptic flow, $v_2$, we use PHOBOS data for Au-Au
\cite{Back:2004mh} and Cu-Cu~\cite{Alver:2006wh} collisions. The same
analysis could be carried out using data from
PHENIX~\cite{Adare:2006ti} or STAR~\cite{Adams:2004bi}.  The initial
eccentricity $\varepsilon$ and the transverse density $(1/S)(dN/dy)$
are evaluated using a model of the collision.  Two such models will be
compared.  The remaining parameters $v_2^{\rm hydro}$ and $\sigma$ are
fit to the data. The first step is to plot $v_2/\varepsilon$ versus
$(1/S)(dN/dy)$~\cite{Voloshin:1999gs}.  Such plots have already been
obtained at SPS and RHIC~\cite{Alt:2003ab}, and they are puzzling:
while $v_2/\varepsilon$ increases with centrality, it shows no hint of
the {\em saturation} predicted by Eq.~(\ref{v2k}) for $K/K_0\simle 1$,
suggesting that the system is far from
equilibrium~\cite{Bhalerao:2005mm}.  On the other hand, the value of
$v_2$ for central Au-Au collisions at RHIC is about as high as
predicted by hydrodynamics, which is widely considered as key evidence
that a ``perfect liquid'' has been created at
RHIC~\cite{Tannenbaum:2006ch}.

It was understood only recently that the eccentricity of the overlap
zone has so far been underestimated, as the result of two effects.
The first effect is fluctuations in initial
conditions~\cite{Socolowski:2004hw}: the time scale of the
nucleus-nucleus collision at RHIC is so short that each nucleus
remains in a frozen configuration, with its nucleons distributed
according to the nuclear wave function.  Fluctuations in the nucleon
positions result in fluctuations of the overlap area. Their effect on
elliptic flow was first pointed out in Ref.~\cite{Miller:2003kd}. It
was later realized by the PHOBOS
collaboration~\cite{Alver:2006wh,Manly:2005zy} that the orientation of
the almond may also fluctuate, so that $\Phi_R$ in Eq.~(\ref{defv2})
is no longer the direction of impact parameter, but the minor axis of
the ellipse defined by the positions of the nucleons. These
fluctuations explain both the large magnitude of $v_2$ for small
systems, such as Cu-Cu collisions, as well as the non-zero magnitude
of $v_2$ in central collisions, where the eccentricity would otherwise
vanish. They have to be taken into account in order to observe the
expected saturation of $v_2/\varepsilon$ at high density mentioned
above.

The eccentricity is usually estimated from the distribution of
participant nucleons in the transverse plane (Glauber model). More
precisely, we assume here that the density distribution of produced
particles is given by a fixed 80\%:20\% superposition of participant
and binary-collision scaling, respectively~\cite{Kharzeev:2000ph}.
For Au-Au collisions, this simple model reproduces the centrality
dependence of the multiplicity reasonably well (we assume that charged
particles are 2/3 of the total multiplicity, and that
$dN/d\eta\simeq 0.8\, dN/dy$ at midrapidity), while it
underestimates it for central Cu-Cu collisions by about 10\%.

At high energies a second effect which increases the
eccentricity is perturbative gluon saturation, which determines the
$p_\perp$-integrated multiplicity from weak-coupling QCD without
additional models for soft particle production.  High-density QCD (the
``Color-Glass Condensate'') predicts a different distribution of
produced gluons, $dN/d^2{\bf r}_\perp dy$, which gives a similar
centrality dependence of the multiplicity~\cite{Kharzeev:2000ph} but a
larger eccentricity~\cite{Hirano:2005xf,Drescher:2006pi}.  When
particle production is dominated by transverse momenta below the
saturation scale of the denser nucleus, then $dN/d^2{\bf r}_\perp
dy\sim {\rm min}(n^A_{\rm part}({\bf r}_\perp), n^B_{\rm part}({\bf
r}_\perp))$ traces the participant density of the more dilute
collision partner, rather than the average as in the Glauber
model~\cite{Drescher:2006pi}. Precise figures depend on how the
saturation scale is defined~\cite{Lappi:2006xc}. Naively, the larger
initial eccentricity predicted by the gluon saturation approach is
expected to require more dissipation in order to reproduce the same
experimentally measured $v_2$. Somewhat surprisingly, we shall find
that this expectation is incorrect, which underscores the non-trivial
role played by the initial conditions.

Both effects, fluctuations and gluon saturation, were recently
combined by Drescher and Nara~\cite{Drescher:2006ca}. In their
approach, the saturation momenta and the unintegrated gluon
distribution functions of the colliding nuclei are determined for each
configuration individually. The finite interaction range of the
nucleons is also taken into account. Upon convolution of the
projectile and target unintegrated gluon distribution functions and
averaging over configurations, the model leads to a very good
description of the multiplicity for both Au-Au as well as Cu-Cu
collisions over the entire available range of centralities.

Having determined the density distributions of produced particles from
either model as described above, we obtain the eccentricity
via~\cite{Miller:2003kd,Bhalerao:2006tp}
\begin{equation}
\varepsilon = \sqrt{\langle\varepsilon_{\rm
    part}^2\rangle}\quad,\quad
\varepsilon_{\rm part} = \frac{\sqrt{(\sigma_y^2-\sigma_x^2)^2 +
    4\sigma_{xy}^2}}
{\sigma_x^2+\sigma_y^2}~.
\end{equation}
$\sigma_x$, $\sigma_y$ are the respective root-mean-square widths of
the density distributions, and $\sigma_{xy}=\overline{xy}-\bar{x}
\bar{y}$ (a bar denotes a convolution with the density distribution
for a given configuration while brackets stand for averages over
configurations). The overlap area $S$ is defined by $S\equiv
4\pi\sigma_x\sigma_y$~\cite{Gombeaud:2007ub}. We find it more
appropriate to define these moments via the number density
distribution $dN/d^2{\bf r}_\perp dy$ rather than the energy density
distribution $dE_\perp/d^2{\bf r}_\perp dy$. 
The reason is twofold: first, $v_2$ is extracted experimentally from
the azimuthal distribution of particle number, not transverse energy;
second, our CGC approach describes the centrality dependence of the
{\em measured} final-state multiplicity very well, which indicates
that the ratio of final-state particles to initial-state gluons
(including possible gluon multiplication processes~\cite{BMSS}) is
essentially constant.

\begin{figure}
\includegraphics*[width=\linewidth]{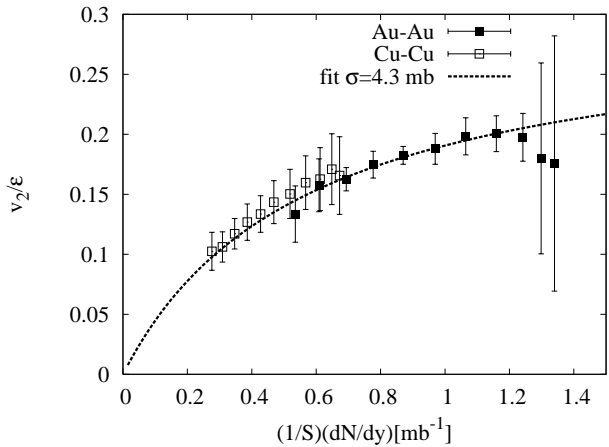}
\caption{Variation of the scaled elliptic flow with the density,
  assuming initial conditions from the Glauber model. 
The line is a 2-parameter fit using Eqs.~(\ref{v2k}) and
  (\ref{knud}). 
\label{fig:glauber}}
\end{figure}

\begin{figure}
\includegraphics*[width=\linewidth]{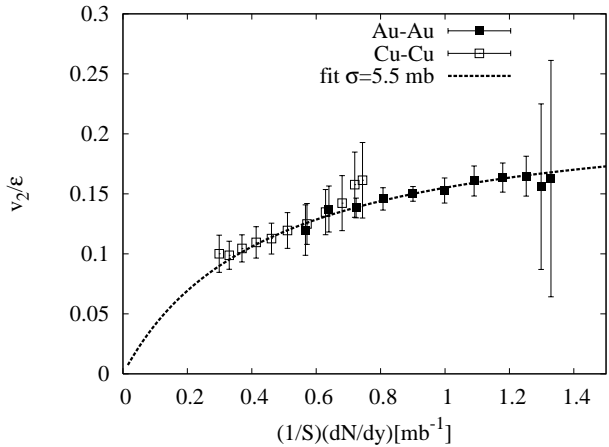}
\caption{Same as Fig.~\ref{fig:glauber}, using CGC initial conditions.
\label{fig:cgc}}
\end{figure}

Figs.~\ref{fig:glauber} and \ref{fig:cgc} display $v_2/\varepsilon$ as
a function of $(1/S)(dN/dy)$ for Au-Au and Cu-Cu collisions at various
centralities, within the Glauber and CGC approaches, respectively. For
both types of initial conditions, Cu-Cu and Au-Au collisions at the
same $(1/S)(dN/dy)$ give the same $v_2/\varepsilon$ within error bars.
Eccentricity fluctuations are crucial for this
agreement~\cite{Alver:2006wh}.  The figures also show that
Eqs.~(\ref{v2k}) and (\ref{knud}) provide a good fit to the data, for
both sets of initial conditions.  On the other hand, the values of the
fit parameters clearly depend on the initial conditions, which has
important consequences for the physics.

The first physical quantity extracted from the fit is the hydrodynamic
limit, $v_2^{\rm hydro}/\varepsilon$, obtained by extrapolating to
$(1/S)(dN/dy)\to \infty$.  The values are $v_2^{\rm hydro} /
\varepsilon = 0.30\pm 0.02$ with the Glauber parameterization, and
$v_2^{\rm hydro}/\varepsilon=0.22\pm 0.01$ with CGC initial
conditions. Comparing these numbers to the experimental data points
one observes that deviations from ideal hydrodynamics are as
large as 30\%, even for central Au-Au collisions. This is our first
important result.

So far, a quantitative extraction of the QCD EoS from
RHIC data via hydrodynamic analysis was hampered by the fact that
$v_2/\varepsilon$ had not been factorized into the perfect-fluid part
$v_2^{\rm hydro}/\varepsilon$ and the dissipative correction
$1/(1+K/K_0)$.  For example, Huovinen found~\cite{Huovinen:2005gy}
that an EoS with a rapid cross-over rather than a strong first-order
phase transition, as favored by lattice QCD~\cite{Bernard:2006nj},
overpredicted the flow data. This finding was rather puzzling, too, as
it was widely believed that the RHIC data fully saturates the
hydrodynamic limit. Our results suggest that ideal hydrodynamics {\em
should} in fact overpredict the measured flow. That is, that one should
not choose an EoS in perfect-fluid simulations that fits the
data. Rather, the EoS could be extracted by comparing ideal
hydrodynamics to $v_2^{\rm hydro}/\varepsilon$.

The next result is that CGC initial conditions, which predict a higher
initial eccentricity $\varepsilon$, naturally lead to a lower
hydrodynamic limit $v_2^{\rm hydro}/\varepsilon$.  Now, close to the
ideal-gas limit ($c_s=1/\sqrt{3}$), $v_2^{\rm hydro}/\varepsilon$
scales approximately like the sound velocity
$c_s$~\cite{Bhalerao:2005mm}.  This means that CGC initial conditions
imply a lower average speed of sound (softer equation of state) than
Glauber initial conditions, by a factor of $0.22/0.3\simeq 0.73$.

The second fit parameter is the partonic cross section $\sigma$. The
larger $\sigma$, the faster the saturation of $v_2/\varepsilon$ as a
function of $(1/S)(dN/dy)$. For our standard values of $K_0$ and $c_s$
we obtain $\sigma=4.3\pm 0.6$~mb for Glauber initial conditions and
$\sigma=5.5\pm 0.5$~mb for CGC initial conditions. These values are
significantly smaller than those found in previous transport
calculations~\cite{Molnar}, but match the findings of
ref.~\cite{XuGreiner}.

CGC initial conditions imply a larger value of $\sigma$ than Glauber
initial conditions, that is, a {\it lower} viscosity.  This can be
easily understood.  As already mentioned above, the CGC predicts a
larger eccentricity $\varepsilon$ than the Glauber model for
semi-central collisions of large nuclei (when there is a large
asymmetry in the local saturation scales of the collision partners,
along a path in impact-parameter direction away from the
origin~\cite{Drescher:2006pi}). However, for very peripheral
collisions or small nuclei, there is of course very little asymmetry
in the saturation scales, and the eccentricity approaches the same
value as in the Glauber model. This has been checked numerically in
fig.~7 of ref.~\cite{Drescher:2006ca}, and can also be clearly seen by
comparing our figures: while in Fig.~\ref{fig:cgc} $v_2/\varepsilon$
for semi-central Au+Au collisions is lower than in
Fig.~\ref{fig:glauber}, there is no visible difference for peripheral
Cu+Cu collisions. In all, with CGC initial conditions the scaled flow
grows less rapidly with the transverse density, which is the reason
for the larger elementary cross-section.

The dependence of $\sigma$ on the initial conditions is probably even
stronger than the numerical values above suggest, for the following
reason.  As alluded to above, our fit to the data really determines
the product $K_0\sigma c_s$, rather than $\sigma$ alone.  It appears
reasonable to assume that $K_0$ does not depend on the initial
conditions. However, for consistency, the speed of sound $c_s$
entering the Knudsen number should match the one underlying the
hydrodynamic limit $v_2^{\rm hydro} / \varepsilon$; hence, if CGC
initial conditions require a smaller $c_s$ by a factor $0.73$, the
elementary cross-section obtained above should be rescaled
accordingly. This leads to our final estimate $\sigma_{CGC}\simeq
7.6\pm 0.7$~mb.

Our numerical results for $\sigma$ should be taken as rough estimates
rather than precise figures, because of the uncertainties related to
the precise values of $K_0$ and $c_s$. It is, however, tempting
to convert them into estimates of the shear viscosity 
$\eta$, which has been of great interest lately.  A
universal lower bound $\eta/s\ge 1/4\pi$ (where $s$ is the entropy
density) has been conjectured using a correspondence with black-hole
physics~\cite{Kovtun:2004de}, and it has been argued that the
viscosity of QCD might be close to the lower bound.  Extrapolations of
perturbative estimates to temperatures $T\simeq200$~MeV, on the other
hand, suggest that the viscosity of QCD could be much
larger~\cite{Huot:2006ys}.  On the microscopic side, $\eta$ is related
to the scattering cross-section $\sigma$. Following
Teaney~\cite{Teaney:2003kp}, the relation for a classical gas of
massless particles with isotropic differential cross sections 
(which applies, for example, to a Boltzmann-transport model)
is $\eta=1.264\, T/\sigma$~\cite{hardspheres}. On the other hand, the
entropy density of a classical ultrarelativistic cas is
$s=4n$, with $n$ the particle density, so that
\begin{equation}
\label{eta}
\frac{\eta}{s}=0.316\frac{T}{c\sigma n}=0.316\frac{\lambda T}{c}.
\end{equation}
The relevant particle density in Au-Au collisions at RHIC, which is
estimated at the time when $v_2$ develops~\cite{Bhalerao:2005mm}, is
3.9~fm$^{-3}$ for both Glauber and CGC initial conditions, and
$T\simeq 200$~MeV. Our two estimates $\sigma=4.3$~mb (Glauber initial
conditions) and $\sigma=7.6$~mb (CGC initial conditions) thus
translate into $\lambda=0.60$~fm, $\eta/s=0.19$ and $\lambda=0.34$~fm,
$\eta/s=0.11$, respectively.  These values for $\eta/s$ agree with
those from ref.~\cite{lacey} if the mean-free path is scaled to our
result, and also with estimates of $\eta/s$ based on the observed
energy loss and elliptic flow of heavy quarks~\cite{phenixQ}, on
transverse momentum correlations~\cite{GavinAziz}, or
bounds on entropy production~\cite{DMN}. Hence,
for our best fit(s) $\eta/s$ is slightly larger than the conjectured
lower bound, but significantly smaller than extrapolations from
perturbative estimates. On the other hand, our lower value is close to
a recent lattice estimate~\cite{Meyer:2007ic} for SU(3) gluodynamics,
which gives $\eta/s=0.134\pm 0.033$ at $T=1.65\, T_c$.

A complementary approach to incorporate corrections from the
ideal-fluid limit is viscous relativistic hydrodynamics. A formulation
that is suitable for applications to high-energy heavy-ion collisions
has been developped in recent years~\cite{Muronga:2001gn}.  A first
calculation of elliptic flow~\cite{Romatschke:2007mq} shows that for 
Glauber initial
conditions and $\eta/s=0.16$, $v_2$ reaches about $70\%$ of the
ideal-fluid value for semi-central Au-Au collisions. It is interesting
to note that our simple estimates are in good agreement with this
finding.  Using Eq.~(\ref{eta}), $\eta/s=0.16$ corresponds to
$\sigma=5.1$~mb, for which Eqs.~(\ref{v2k}) and (\ref{knud}) give
$v_2/v_2^{\rm hydro}=0.68$.  The comparison to experimental data in
Ref.~\cite{Romatschke:2007mq}, however, appears to favor lower values
of $\eta/s$ because the EoS used there underpredicts $v_2^{\rm
hydro}/\varepsilon\simeq0.3$ required for Glauber initial
conditions. Alternatively, simulations could be performed with CGC
initial conditions which require only $v_2^{\rm
hydro}/\varepsilon\simeq0.22$.

In summary, we have shown that the centrality and system-size
dependence of the {\em measured} $v_2$ can be understood as follows:
$v_2$ scales like the initial eccentricity $\varepsilon$ (as predicted
by hydrodynamics), multiplied by a correction factor due to
off-equilibrium (i.e., viscous) effects. This correction involves the
multiplicity density in the overlap area, $(1/S)(dN/dy)$.  Two types
of initial conditions have been compared: a Glauber-type model, and a
Color-Glass Condensate approach. PHOBOS data can be described with
both. In particular, there is good agreement between Cu-Cu and Au-Au
data. The resulting estimates for thermodynamic quantities and
transport coefficients, on the other hand, depend significantly on the
initial conditions.

Color glass condensate-type initial conditions require {\em lower}
viscosity and a {\em softer} equation of state (smaller speed of
sound). The scaled flow extrapolated to vanishing mean-free path is
lower than for Glauber initial conditions by a factor of $\simeq
0.22/0.3=0.73$; the effective speed of sound should also be lower by
about the same factor. Our estimates for the viscosity are
$\eta/s\simeq 0.19$ for Glauber initial conditions, and $\eta/s\simeq
0.11$ for CGC initial conditions, but these numbers should be taken
only as rough estimates.

We have also shown that the data for the scaled flow indeed {\em
saturate} at high densities to a hydrodynamic limit. In central Au-Au
collisions at RHIC, $v_2$ reaches 70\% (resp.\ 75\%) of the
hydrodynamic limit for Glauber (CGC) initial conditions. The
corrections to ideal hydrodynamics are therefore significant, but
reasonably small compared to unity, implying that (viscous)
hydrodynamics should be a valid approach for understanding flow at
RHIC. Also, the asymptotic limit of $v_2/\varepsilon$ has been
isolated and could now be used to test realistic equations of state
from lattice-QCD with hydrodynamic simulations of heavy-ion
collisions.

\section*{Acknowledgments}

J.Y.O.\ thanks B.\ Alver, A.H.\ Mueller and D.\ Schiff for helpful
discussions.  H.J.D.\ is supported through BMBF grant 05 CU5RI1/3.

\end{document}